\documentclass[prl,twocolumn,showpacs,amsmath,amssymb,aps,superscriptaddress]{revtex4}
\usepackage{graphicx}
\usepackage{dcolumn}
\usepackage{bm}
\usepackage{txfonts}
\usepackage{amsmath}
\usepackage{epsf}
\usepackage{epsfig}
\usepackage{amssymb}
\usepackage{color}
\usepackage{multirow}
\usepackage{pst-3d}
\usepackage{rotating}

\begin{document}

\title{Supercooling of Atoms in an Optical Resonator}

\author{Minghui Xu}
\affiliation{JILA, National Institute of Standards and Technology and
  Department of Physics, University of Colorado, Boulder, Colorado
  80309-0440, USA}
\author{Simon B. J\"ager}
\affiliation{Theoretische Physik, Universit\"at des Saarlandes, 
D-66123 Saarbr\"ucken, Germany}
\author{S. Sch\"utz}
\affiliation{Theoretische Physik, Universit\"at des Saarlandes, 
D-66123 Saarbr\"ucken, Germany}
\author{J. Cooper}
\affiliation{JILA, National Institute of Standards and Technology and
  Department of Physics, University of Colorado, Boulder, Colorado
  80309-0440, USA}
\author{Giovanna Morigi}
\affiliation{Theoretische Physik, Universit\"at des Saarlandes, 
D-66123 Saarbr\"ucken, Germany}
\author{M. J. Holland}
\affiliation{JILA, National Institute of Standards and Technology and
  Department of Physics, University of Colorado, Boulder, Colorado
  80309-0440, USA}
\date{\today}%

\begin{abstract}
  We investigate laser cooling of an ensemble of atoms in an optical
  cavity.  We demonstrate that when atomic dipoles are sychronized in
  the regime of steady-state superradiance, the motion of the atoms
  may be subject to a giant frictional force leading to potentially
  very low temperatures.  The ultimate temperature limits are
  determined by a modified atomic linewidth, which can be orders of
  magnitude smaller than the cavity linewidth.  The cooling rate is
  enhanced by the superradiant emission into the cavity mode allowing
  reasonable cooling rates even for dipolar transitions with
  ultranarrow linewidth.
\end{abstract}


\maketitle

The discovery of laser cooling~\cite{Wineland99} has enabled a new
world of quantum gas physics and quantum state engineering in dilute
atomic systems~\cite {Bloch08}. Laser cooling is an essential
technology in many fields, including precision measurements, quantum
optics, and quantum information
processing~\cite{Jun15,Walmsley15,Wineland13}.  Doppler
cooling~\cite{Schawlow75,Wineland79} is perhaps the most elementary
kind of laser cooling and relies on repeated cycles of electronic
excitation by lasers followed by spontaneous relaxation. The
temperatures that can be achieved in this way are limited by the
atomic linewidth.  Only specific ionic and atomic species can be
Doppler cooled because they typically should possess an internal level
structure that allows for closed cycling transitions.

Cavity-assisted laser cooling~\cite{Ritsch03,Ritsch13} utilizes the
decay of an optical resonator instead of atomic spontaneous emission
as the energy dissipation mechanism.  It is based on the preferential
coherent scattering of laser photons into an optical cavity~\cite
{Ritsch97,Chu00}, rather than absorption of free-space laser photons
as in conventional Doppler cooling. Temperatures that can be achieved
in cavity-assisted cooling are limited by the cavity linewidth. Since
the particle properties enter only through the coherent scattering
amplitude, cavity-assisted cooling promises to be applicable to any
polarizable object~\cite{Kimble03,Rempe04,Vuletic09,Vuletic11,
  Hemmerich13,Morigi07,Ye08,Aspelmeyer13,Barker15}, including
molecules~\cite{Morigi07,Ye08} and even mesoscopic systems such as
nanoparticles~\cite{Aspelmeyer13,Barker15}.

The many-atom effects of cavity-assisted cooling were theoretically
discussed by Ritsch and collaborators~\cite{Ritsch02} and
experimentally reported in Refs.~\cite{Vuletic03,Barrett12}.  The
cavity-mediated atom-atom coupling typically leads to a cooling rate
that is faster for an atomic ensemble than for a single atom. Above a
threshold of the pump laser, self-organization may occur and is
observed as patterns in the atomic distribution that maximize the
cooperative scattering.  Recently, it has been shown that the
long-range nature of the cavity-mediated interaction between atoms
gives rise to interesting prethermalization behavior in the
self-organization dynamics~\cite{Morigi14}. In spite of the intrinsic
many-body nature, the underlying cooling mechanism shares much with
the single-atom case, and indeed the final temperature observed in
these systems is limited by the cavity linewidth.

In this paper, we demonstrate that the mechanical action of the
atom-cavity coupling takes on a dramatically new character for atoms
in the regime of steady-state superradiance~\cite{Meiser09,Meiser10,
  Thompson12,Thompson13,Xu14,Xu15}. Specifically, the frictional force
on a single atom is significantly enhanced, and the final temperature
is much lower than the temperature that can be achieved in
cavity-assisted cooling~\cite{Ritsch97,Chu00}.  Furthermore, as the
atom number increases, the cooling may become faster due to the
increasing rate of superradiant collective emission. We show that
ability to achieve much lower temperatures than for single-atom
cavity-assisted cooling derives from the emergence of atom-atom dipole
correlations in the many-body atomic ensemble.

\begin{figure}[b]
  \centerline{\includegraphics[width=0.7\linewidth, angle=0]{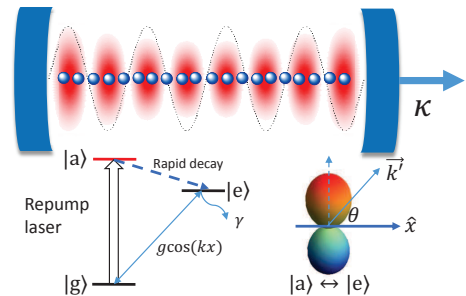}}
  \caption{\label{Fig1}(color online) Atoms with ultranarrow
    transition $\bigl|g\bigr>\leftrightarrow\bigl|e\bigr>$ are
    confined to the axis of a standing-wave mode of an optical
    cavity. Different implementations of pumping may be
    considered~\cite{Meiser09,Thompson12}. In the simplest scenario
    shown, a transition is driven from the ground state $|g\rangle$ to
    an auxiliary state $|a\rangle$ that rapidly decays to the excited
    state~$|e\rangle$. In this way $|a\rangle$ can be adiabatically
    eliminated and a two-state pseudospin description in the
    $\{|g\rangle,|e\rangle\}$ subspace used, with repumping
    corresponding to an effective rate $w$ from $|g\rangle$ to
    $|e\rangle$. If the repumping laser is directed normal to the
    cavity axis, the absorption does not modify the momentum. Momentum
    recoil is induced by the on-axis component of the wavevector
    $\vec{k'}$ of the dipole radiation pattern for the
    $|a\rangle\leftrightarrow|e\rangle$ transition. }
\end{figure}

Steady-state superradiant lasers were proposed in Ref.~\cite{Meiser09}
as possible systems for generating millihertz linewidth light, and
demonstrated in a recent experiment using a two-photon Raman
transition~\cite{Thompson12}. In the regime of steady-state
superradiance, the cavity decay is much faster than all other
processes. Therefore, the cavity mode plays the role of a dissipative
collective coupling for the atoms that leads to the synchronization of
atomic dipoles~\cite{Xu14,Xu15}.  The emergence of a macroscopic
collective dipole induces an extremely narrow linewidth for the
generated light~\cite{Meiser09,Xu15}. The optimal parameters are in
the weak-coupling regime of cavity QED~\cite{Book}, that is opposite
to the strong-coupling situation usually considered in cavity-assisted
cooling~\cite{Ritsch03,Ritsch13}. Superradiant lasers require
weak-dipole atoms~({\it e.g.}\ using intercombination lines or other
forbidden transitions) confined in a high-finesse optical cavity.

With this background, we now consider a specific situation of an
ensemble of $N$ point-like two-level atoms with transition frequency
$\omega_\mathrm{a}$ and natural linewidth $\gamma$, interacting with a
single-mode cavity with resonance frequency $\omega_\mathrm{c}$ and
linewidth $\kappa$, as shown in Fig.~\ref{Fig1}.  The atoms are
restricted to move freely along the direction of the cavity
axis~($x$-axis), a situation that can be realized by tightly confining
the atoms in the other two directions. The atom-cavity coupling
strength is given by $g\cos(kx)$, where $g$ is the vacuum Rabi
frequency at the field maximum and $\cos(kx)$ describes the
one-dimensional cavity mode function.  The atoms are incoherently
repumped at rate $w$, thus providing the source of photons.

The Hamiltonian describing the atom-cavity system in the rotating
frame of the atomic transition frequency is given by,
\begin{equation}
\hat{H}=\hbar\Delta\hat{a}^\dagger\hat{a}+
\sum_{j=1}^N\frac{\hat{p}_j^2}{2m}
+\hbar\frac{g}{2}\sum_{j=1}^{N}(\hat{a}^\dagger
\hat{\sigma}_j^-+\hat{\sigma}_j^+\hat{a})\cos(k\hat{x}_j)\,,
\end{equation}
where $\Delta=\omega_\mathrm{c}-\omega_\mathrm{a}$.  We have
introduced the bosonic annihilation and creation operators, $\hat{a}$
and $\hat{a}^\dagger$, for cavity photons. The $j$-th atom is
represented by Pauli pseudospin operators, $\hat{\sigma}_j^z$ and
$\hat{\sigma}_j^-=(\hat{\sigma}_j^+)^\dagger$, and position and
momentum $\hat{x}_j$ and $\hat{p}_j$, respectively.

In the presence of dissipation, the evolution of the system is
described by the Born-Markov quantum master equation for the density
matrix $\hat{\rho}$ for the cavity and atoms,
\begin{equation}\label{masterEq}
\frac{d}{dt}\hat{\rho}=\frac{1}{i\hbar}\left[\hat{H},\hat{\rho}\right]
+\kappa\mathcal{L}[\hat{a}]\rho+w
\sum_{j=1}^N\int_{-1}^{1} du N(u)
\mathcal{L}[\hat{\sigma}_j^+ e^{iuk'\hat{x}_j}]\rho\,,
\end{equation}
where
$\mathcal{L}[\hat{\mathcal{O}}]\hat{\rho}=(2\hat{\mathcal{O}}\hat{\rho}
\hat{\mathcal{O}}^\dagger-\hat{\mathcal{O}}^\dagger\hat{\mathcal{O}}\hat{\rho}
-\hat{\rho}\hat{\mathcal{O}}^\dagger\hat{\mathcal{O}})/2$ is the
Linbladian superoperator describing the incoherent processes. The term
 proportional to $\kappa$ describes the cavity decay. The repumping is
the term proportional to $w$ and is modeled by spontaneous absorption
with recoil~\cite{Book2}. The recoil is parametrized by the normalized
emission pattern $N(u)$ and wavevector $k'$.  It will generally be a
good approximation to neglect the effect of free-space spontaneous
emission of the atoms out the side of the cavity, since the natural
linewidth $\gamma$ is assumed to be extremely small for atoms with an
ultraweak-dipole transition.

In the parameter regime of interest, the cavity linewidth is much
larger than other system frequencies, and the cavity field can be
adiabatically eliminated, resulting in the phase locking of the cavity
field to the collective atomic dipole~\cite{Meiser10,Xu14,Xu15}. In
order to correctly encapsulate the cavity cooling mechanism, the
adiabatic elimination of the cavity field has to be expanded beyond
the leading order. Specifically, the retardation effects between the
cavity field and atomic variables should be included.  As shown in the
Supplemental Material~\cite{suppl}, in the large $\kappa$
limit~\cite{note},
\begin{equation}\label{ada}
  \hat{a}(t)\approx\frac{-i\frac{g}{2}\hat{J}^-}{\kappa/2+i\Delta}
  +\frac{\frac{d}{dt}(i\frac{g}{2}\hat{J}^-)}{(\kappa/2+i\Delta)^2}
  -\frac{2i\sqrt{\Gamma_C}}{g}\hat{\xi}(t)
  +\mathcal{O}[\kappa^{-3}]\,,
\end{equation}
where $\hat{J}^-=\sum_{j=1}^N\hat{\sigma}_j^-\cos(k\hat{x}_j)$ is the
collective dipole operator,
$\Gamma_C=g^2\kappa/4(\kappa^2/4+\Delta^2)$ is the atomic spontaneous
emission rate through the cavity, and $\hat{\xi}(t)$ is the quantum
noise originating from the vacuum field entering through the cavity
output.

The dipole force on the $j$-th atom is given by the gradient of the
potential energy, which takes the form
\begin{equation}
\label{dipoleforce}
F_j=\frac{d}{dt}\hat{p}_j=-\nabla_j\hat{H}=\frac12\hbar k g
\sin(k\hat{x}_j)
\left(\hat{\sigma}_j^+\hat{a}+\hat{a}^\dagger\hat{\sigma}_j^-\right)\,.
\end{equation}
We maximize the single-atom dissipative force by working at the
detuning $\Delta=\kappa/2$~\cite{suppl}, and in that case by
substituting Eq.~(\ref{ada}) into Eq.~(\ref{dipoleforce}) we find
\begin{equation}\label{force}
\begin{split}
\frac{d}{dt}&\hat{p}_j\approx
{}-\frac12\hbar k\Gamma_C\sin(k\hat{x}_j)\biggl((1+i)
\hat{\sigma}_j^+\hat{J}^-
+(1-i)\hat{J}^+\hat{\sigma}_j^-\biggr)\\
&{}-\frac12\eta\Gamma_C\sin(k\hat{x}_j)\sum_{l=1}^N
(\hat{\sigma}_j^+\hat{\sigma}_l^-+\hat{\sigma}_l^+\hat{\sigma}_j^-)
\frac{1}{2} [\sin(k\hat{x}_l),\hat{p}_l]_+
+\hat{\mathcal{N}}_j\,.
\end{split}
\end{equation}
Here the anticommutator is
$[ \hat{A}, \hat{B}]_+ =\hat{A} \hat{B} + \hat{B} \hat{A} $. We have
also defined $\eta=4\omega_\mathrm{r}/\kappa$, which characterizes the
likelihood of a photon emission into the cavity mode in the direction
of motion versus the opposite, in terms of the atomic recoil frequency
$\omega_\mathrm{r}=\hbar k^2/2m$.  The three terms on the right hand
side of Eq.~(\ref{force}) can be interpreted as the conservative
force, the friction, and the noise-induced momentum fluctuations,
respectively.

At temperatures much greater than the recoil temperature the motion is
well described by a semiclassical treatment. A systematic
semiclassical approximation, to make the mapping
$\langle\hat{x}_j\rangle\rightarrow x_j$ and
$\langle\hat{p}_j\rangle\rightarrow p_j$ where $x_j$ and $p_j$ are
classical variables, is based on the symmetric ordering of operator
expectation values. In order to accurately incorporate the effects of
quantum noise, we match the equations of motion for the second-order
moments of momenta between the quantum and semiclassical theories so
that we obtain the correct momentum diffusion~\cite{suppl}. This
procedure yields Ito stochastic equations,
\begin{equation}\label{peom}
\begin{split}
\frac{d}{dt}p_j\approx&
\hbar k\Gamma_C\sin(kx_j)\left(\mathrm{Im}[\langle\hat{\sigma}_j^+
\hat{J}^-\rangle]
-\mathrm{Re}[\langle\hat{\sigma}_j^+\hat{J}^-\rangle]\right)\\
&{}-\eta\Gamma_C\sin(kx_j)\sum_{l=1}^N
\mathrm{Re}[\langle\hat{\sigma}_j^+\hat{\sigma}_l^-\rangle]
\sin(kx_l)p_l+\xi_j^p\,,
\end{split}
\end{equation}
where $\xi_j^p$ is the classical noise and
$\langle\xi_j^p(t)\xi_l^p(t')\rangle=D^{jl}\delta(t-t')$ with
diffusion matrix
\begin{equation}\label{dm}
\begin{split}
  D^{jl}&=\hbar^2k^2\Gamma_C\sin(kx_j)\sin(kx_l)\mathrm{Re}[\langle
  \hat{\sigma}_l^+\hat{\sigma}_j^-\rangle]\\
  &{}+\hbar^2k'^2w\overline{u^2}
  \langle\hat{\sigma}_j^-\hat{\sigma}_l^+\rangle\delta_{jl}\,,
\end{split}
\end{equation}
involving the geometrical average
$\overline{u^2}\equiv\int_{-1}^1u^2N(u)\,du$ and Kronecker delta
$\delta_{jl}$. The momentum evolution is paired with the usual
equation for $x_j$
\begin{equation}
\label{xeom}
\frac{d}{dt}x_j=\frac{p_j}{m}\,.
\end{equation}

\begin{figure}[t]
  \centerline{\includegraphics[width=0.7\linewidth, angle=0]{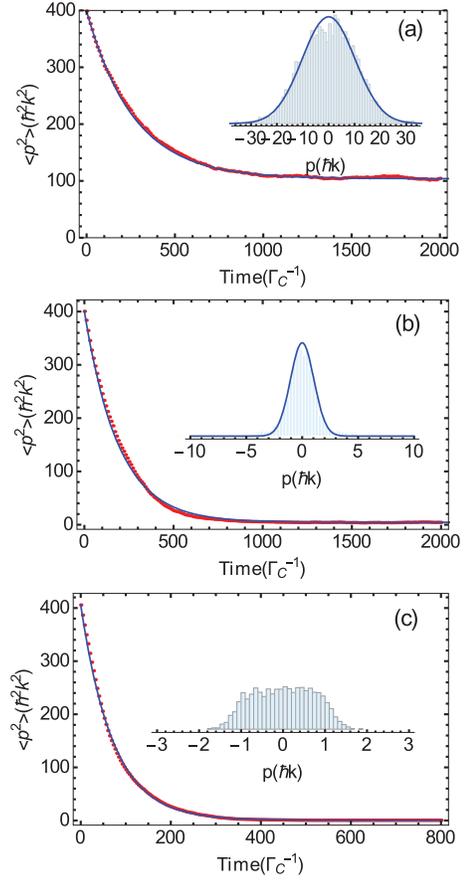}}
  \caption{\label{Fig2}(color online) Time evolution of the average
    momentum square~(red dots) evaluated from 4000 trajectories
    simulated by integrating Eqs.~(\ref{peom}) and (\ref{xeom}) for 1
    atom~(a), 20 atoms~(b), and 60 atoms~(c).  The blue solid line is
    a fit to an exponential decay. The parameters are
    $\Delta=\kappa/2=100$, $\Gamma_C=0.1$, and
    $\omega_\mathrm{r}=0.25$. The repumping rates are chosen such that
    the average atomic population inversion in all cases is the
    same~[$w=0.15$~(a), 0.28~(b), 1.3~(c)].  Insets show the momentum
    statistics. The blue solid line is a fit to a Gaussian
    distribution.}
\end{figure}

To begin with, we first consider the case in which the effect of
recoil associated with the repumping is neglected, {\em i.e.}\ we set
$k'=0$. This will determine the ultimate temperature limit imposed by
the vacuum noise due to the cavity output. For the simple one-atom
case, we can then directly find the friction~($\alpha$) and
diffusion~($D$) coefficient from Eq.~(\ref{peom}) and
Eq.~(\ref{dm}). The steady-state temperature~$T$ for the single atom
(labeled by 1) is
\begin{equation}\label{oac}
k_BT=\frac{\langle p_1^2\rangle}{m}
=\frac{D}{2m\alpha}={\hbar\kappa\over4}\,,
\end{equation}
since 
\begin{eqnarray}
D&=&\hbar^2k^2\Gamma_C\sin^2(kx_1)
  \langle\hat{\sigma}_1^+\hat{\sigma}_1^-\rangle\,,\nonumber\\
\alpha&=&\eta\Gamma_C\sin^2(kx_1)
  \langle\hat{\sigma}_1^+\hat{\sigma}_1^-\rangle\,.
\end{eqnarray}
Note that this is precisely the same temperature limit previously
found in the cavity-assisted cooling case where the system is
operating in the strong coupling cavity-QED region. Here the rate of
the decay into the cavity mode is proportional to
$\Gamma_C\langle\hat{\sigma}_1^+\hat{\sigma}_1^-\rangle$, which is
applicable to the weak coupling regime of cavity QED~\cite{Book}.  In
Fig.~\ref{Fig2}(a), we show a numerical simulation of the cooling
trajectory of a single atom as a function of time. As expected, the
final temperature $k_B T$ asymptotes to $\hbar\kappa/4$ and the
cooling rate is well approximated by
$R_\mathrm{S}=\eta\Gamma_C\langle\hat{\sigma}_1^+\hat{\sigma}_1^-\rangle$.
 
The cooling in the many-atom case exhibits a distinctly different
character. A feature of this model is the pseudospin-to-motion
coupling of the atoms. In order to close the evolution equations of
the atomic motion as described by Eq.~(\ref{peom}) and
Eq.~(\ref{xeom}), it is also necessary to solve for the dynamics of
the pseudospins. For this purpose, we derive in the Supplemental
Material~\cite{suppl} the effective quantum master equation for the
pseudospins,
\begin{equation}\label{effm}
\frac{d}{dt}\hat{\rho}=\frac{1}{i\hbar}[\hat{H}_\mathrm{eff},\hat{\rho}]
+\Gamma_C\mathcal{L}[\hat{J}^-]\hat{\rho}+w
\sum_{j=1}^N\int_{-1}^{1} du N(u)
\mathcal{L}[\hat{\sigma}_j^+ e^{iuk'\hat{x}_j}]\rho\,,
\end{equation}
where the effective Hamiltonian
$\hat{H}_\mathrm{eff}=-{\hbar\Gamma_C}\hat{J}^+\hat{J}^-/2$
describes the coherent coupling between atoms, and the collective
decay [term proportional to $\Gamma_C$ in Eq.~(\ref{effm})] leads to
dissipative coupling. It is the dissipative coupling that gives rise
to dipole synchronization and steady-state
superradiance~\cite{Meiser09,Meiser10,
  Thompson12,Thompson13,Xu14,Xu15}.
The full pseudospin Hilbert space dimension scales exponentially with
the number of atoms. To solve Eq.~(\ref{effm}), we thus employ a
cumulant approximation that is applicable to large atom
numbers~\cite{Meiser10,Xu14,Xu15}. All nonzero observables are
expanded in terms of $\langle\hat{\sigma}_j^+\hat{\sigma}_j^-\rangle$
and $\langle\hat{\sigma}_j^+\hat{\sigma}_l^-\rangle$~($j\ne l$),
describing the population inversion and spin-spin correlations
respectively.  Their equations of motion are derived in the
Supplemental Material~\cite{suppl}.

\begin{figure}[t]
  \centerline{\includegraphics[width=0.7\linewidth, angle=0]{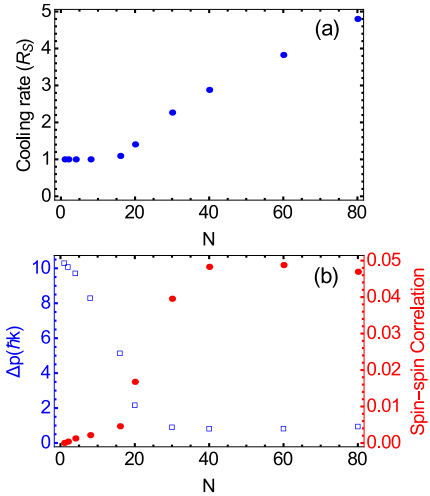}}
  \caption{\label{Fig3}(color online) (a)~Cooling rate~(in units of 
  the single atom cooling rate $R_\mathrm{S}$) as a function of atom number. 
  (b)~Final momentum width~($\Delta p=
  \sqrt{\langle p^2 \rangle}$, blue squares) 
  and spin-spin correlation~(red dots) as a function
  of atom number.  The parameters are the same as those in Fig.~\ref{Fig2}. }
\end{figure}

Simulations of the cooling dynamics for many atoms are shown in
Figs.~\ref{Fig2}(b) and (c). Remarkably, we find the collective atomic
effects to lead to a more rapid cooling rate, and simultaneously to
generate a lower final temperature.  Figure~\ref{Fig3} shows the
 cooling rate (a) and the final momentum width (b) as a function of the atom number.
 We note that the cooling rate exhibits two kinds of behaviour, hinting towards the existence of a $N$-dependent threshold, see Fig.~\ref{Fig3}(a). For $N \lesssim 20 $, the cooling rate is independent of $N$, while for $N \gtrsim 20$, it increases monotonously. Correspondingly, in this regime, the momentum width has reached a minimum which is independent of $N$, see Fig.~\ref{Fig3}(b). Note
that when the final temperature gets closer to the recoil temperature,
the momentum distribution is not Gaussian anymore, rendering the notion
of temperature invalid.  The semiclassical treatment predicts a
uniform distribution in the momentum interval [$-\hbar k$,$\hbar k$]
corresponding to the recoil limit, as shown in the inset of
Fig.~\ref{Fig2}(c). These results demonstrate that for atoms in the
steady-state superradiant regime, not only is the cooling more
efficient due to the rapid rate of superradiant light emission, but
also the final temperature is determined by the relaxation rate of the
atomic dipole, and not the cavity linewidth.

The principal new feature here is that spin-spin correlations between
atoms develop due to the cavity-mediated coupling. In order to measure
the extent of this effect, we introduce
$\langle\hat{\sigma}^+\hat{\sigma}^-\rangle_\mathrm{E}$ defined as
averaged spin-spin correlations,
\begin{equation}
\langle\hat{\sigma}^+\hat{\sigma}^-\rangle_\mathrm{E}
=\left(\langle\hat{J}^+\hat{J}^-\rangle-\sum_{j=1}^N\langle
\hat{\sigma}_j^+\hat{\sigma}_j^-\rangle\cos^2(kx_j)\right)/[N(N-1)]\,.
\end{equation}
Fig.~\ref{Fig3}(b) shows
$\langle\hat{\sigma}^+\hat{\sigma}^- \rangle_\mathrm{E}$ as a function
of the number of atoms.  The equilibrium temperature is seen to
decrease as the collective spin-spin correlation emerges. This is
reminiscent of the linewidth of the superradiant laser, where the
synchronization of spins leads to a significant reduction of the
linewidth to the order of $\Gamma_C$~\cite{Meiser09,Xu15}.  The
establishment of spin-spin correlations is a competition between
dephasing due to both cavity output noise and repumping, and the
dissipative coupling between atoms which tends to synchronize the
dipoles~\cite{Xu15}.  Since the coupling strength scales with $N$, a
sufficient atom number is required to establish strong spin-spin
correlations~\cite{Xu15}.

\begin{figure}[t]
  \centerline{\includegraphics[width=0.7\linewidth, angle=0]{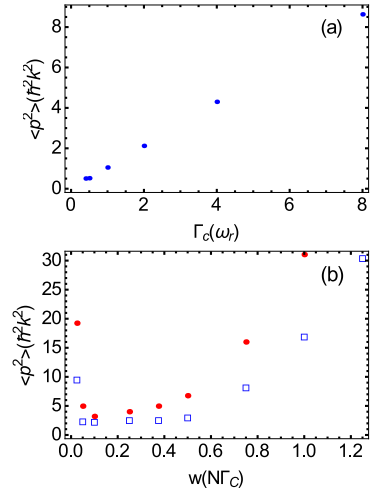}}
  \caption{\label{Fig4}(color online) (a)~Final momentum width as a
    function of $\Gamma_C$ for 40 atoms. The parameters are
    $\Delta=\kappa/2=200$, $w=N\Gamma_C/4$, and
    $\omega_\mathrm{r}=0.25$. (b)~Final momentum width as a function
    of repumping strength for 40 atoms without ($k'=0$, blue squares)
    and with recoil associated with repumping~($k'=k$, red dots).  The
    parameters are $\Delta=\kappa/2=200$, $\Gamma_C=0.5$, and
    $\omega_\mathrm{r}=0.25$. }
\end{figure}

Further characterizing the ultimate temperature limits,
Fig.~\ref{Fig4}(a) shows the final momentum width as a function
of~$\Gamma_C$.  We see that as $\Gamma_C$ is decreased, the final
temperature reduces in proportion to $\Gamma_C$ until it hits the
recoil limit. This effect is consistent with a significantly increased
friction coefficient providing a reduction of the order of the final
temperature from the one to many atom case from $\kappa$ to
$\Gamma_C$.

So far our discussion has neglected the recoil associated with
repumping.  We have done that because its effect on the final
temperature will depend crucially on specifics of its implementation,
including factors such as the polarizations and directions of repump
lasers, the atomic system, and the transitions used. However, in the
specific repumping model shown in Fig.~\ref{Fig1}, the magnitude of
$k'$ controls the recoil effect of the repumping on the momentum
diffusion. Fig.~\ref{Fig4}(b) shows the final momentum width as a
function of repumping for $k'=0$ and $k'=k$. Again, in the region of
small and large repumping, where spin-spin correlations are very
small, the final temperature is high. When the recoil due to repumping
is included, the final temperature becomes higher and is eventually
determined by $w\overline{u^2}$. However for weak repumping, with
$w$ not significantly larger then $\Gamma_C$ it is still possible to
achieve temperatures not much higher than that predicted when pump
recoil was neglected. This is especially promising for the
implementation of supercooling in realistic experimental systems. Note
that $k=k'$ is more or less a worst case senerio, since by using a
dipole allowed transition for the relaxation from the auxiliary state
to the excited state, one could in principle use a much reduced
frequency with correspondingly small recoil.

In conclusion, we have presented a theoretical proposal for
supercooling atoms in one-dimension along the axis of an optical
cavity. The superradiant emission was observed to lead to an increased
cooling rate and to a potentially extremely low final temperature. The
ultimate temperatures were seen to be constrained by the relaxation of
the atomic dipole, and could be orders of magnitude less than the
limits for single atom cooling that are constrained by the cavity
linewidth. This system is an example of many-body laser cooling in
which all motional degrees of freedom of a collective system are
simultaneously cooled and in which macroscopic spin-spin correlations
are essential and must develop for the cooling mechanism to work. It
will be necessary to consider extensions to understand how cooling in
all three dimensions may be achieved, and to consider realistic models
for real atoms that may involve elaborate repumping schemes.

\begin{acknowledgments}
We acknowledge helpful discussions with A. M. Rey, J. Ye, and J. K.
Thompson. This work has been supported by the DARPA QuASAR program,
the NSF, NIST, the German Research Foundation (DACH project ``Quantum
crystals of matter and light''), and the German Ministry of
Education and Research BMBF (Q.Com).
\end{acknowledgments}

\newpage
\setcounter{equation}{0}
\renewcommand*{\citenumfont}[1]{S#1}
\renewcommand*{\bibnumfmt}[1]{[S#1]}
\renewcommand{\thesection}{S.\arabic{section}}
\renewcommand{\thesubsection}{\thesection.\arabic{subsection}}
\makeatletter 
\def\tagform@#1{\maketag@@@{(S\ignorespaces#1\unskip\@@italiccorr)}}
\makeatother
\makeatletter
\makeatletter \renewcommand{\fnum@figure}
{\figurename~S\thefigure}
\makeatother

\onecolumngrid
\begin{center}
\textbf{\large Supplemental Material for\\
Supercooling of Atoms in an Optical Resonator}
\end{center}

\section{I. Adiabatic Elimination of the Cavity Mode}
The regime of steady-state superradiance is defined by a timescale separation between the single cavity mode and the atomic degrees of freedom.
The typical relaxation time of the cavity mode is of the order of $T_C\sim  |\kappa+i\Delta|^{-1}$, while the one of the atoms is given by $T_A\sim \left(\max\left\{\sqrt{N\bar{n}}g,w,k\sqrt{\left\langle p^2\right\rangle}/m\right\}\right)^{-1}$, where $\bar{n}$ is the mean photon number in the cavity.
In order to eliminate the cavity field quasiadiabatically we need the relaxation time of the cavity to be much shorter than the timescale on which the atoms are evolving, namely $T_A\gg T_C$.
To this end, we start with the quantum Langevin equation  
for the cavity field according to the quantum master equation~[Eq.~(2)
in the paper],
\begin{equation}\label{lan}
\frac{d}{dt}\hat{a}=-\frac{\kappa}{2}\hat{a}-i\Delta
\hat{a}-i\frac{g}{2}\hat{J}^-+\sqrt{\kappa}\hat{\xi}(t)\,,
\end{equation}
where $\hat{\xi}(t)$ is the quantum white noise and 
$\langle\hat{\xi}(t)\hat{\xi}^\dagger(t')\rangle
=\delta(t-t')$.
The formal solution to Eq.~(S\ref{lan}) is
\begin{equation}\label{fs}
\hat{a}(t)=e^{-(\kappa/2+i\Delta)\Delta t}\hat{a}(t_0)-i\frac{g}{2}\int_0^{\Delta t}ds
e^{-(\kappa/2+i\Delta)s}\hat{J}^-(t-s)
+\hat{\mathcal{F}}(t)\,,
\end{equation}
where $\hat{\mathcal{F}}(t)=\sqrt{\kappa}\int_0^{\Delta t}dse^{-(\kappa/2+i\Delta)s}\hat{\xi}(t-s)$ is the noise term and $\Delta t=t-t_0$. Under the approximation of coarse
graining~$\left(T_A\gg \Delta t\gg T_C\right)$, the first term on the right-hand side~(RHS) of Eq.~(S\ref{fs})
vanishes, and it can be shown that 
\begin{equation}
\langle\hat{\mathcal{F}}(t)\hat{\mathcal{F}^\dagger}(t')\rangle\approx
e^{-\kappa|t'-t|/2-i\Delta(t-t')}\approx\frac{\kappa}{\kappa^2/4
+\Delta^2}\delta(t-t')\,.
\end{equation}
It would be convenient to choose $\hat{\mathcal{F}}(t)=-i\frac
{\sqrt{\Gamma_C}}{g/2}\hat{\xi}(t)$, with \begin{align}
\Gamma_C=
\frac{g^2\kappa/4}{\kappa^2/4+\Delta^2}.
\end{align} Furthermore, 
the integral in Eq.~(S\ref{fs}) can be expanded in powers of $1/
(\kappa/2+i\Delta)$. As a result we obtain 
\begin{equation}\label{ada}
\hat{a}(t)\approx\frac{-i\frac{g}{2}\hat{J}^-}{\kappa/2+i\Delta}
-\frac{\frac{d}{dt}(-i\frac{g}{2}\hat{J}^-)}{(\kappa/2+i\Delta)^2}
+\hat{\mathcal{F}}(t)+\mathcal{O}[(\kappa/2+i\Delta)^{-3}]\,.
\end{equation}
As can be seen from Eq.~(S\ref{ada}), the retardation effects between
the cavity field and atomic variables are included.\\
\section{II. External motion of atoms}
In this section we derive the force for the external degrees of freedom, including friction and noise. We will end up with a classical description of the particles' external degrees of freedom and derive a Langevin equation for the momenta of the particles. 

The force on the $j$-th atom $\hat{F}_j$ is given by
\begin{equation}
\hat{F}_j=\frac{d}{dt}\hat{p}_j=\hbar
k\sin(k\hat{x}_j)\frac{g}{2}
(\hat{\sigma}_j^+\hat{a}+\hat{a}^\dagger\hat{\sigma}_j^-) + \hat{\mathcal{N}}_j^{\text{pump}}\,,
\end{equation}
where $\hat{\mathcal{N}}_j^{\text{pump}}$
represents the random force due to recoil of the incoherent pumping process.\\ Substituting Eq.~(S\ref{ada}) into the above equation
, we have
\begin{equation}
\begin{split}
\frac{d}{dt}\hat{p}_j &\approx \hbar
k\sin(k\hat{x}_j)\frac{\Gamma_C}{2}\left(-i\hat{\sigma}_j^+\hat{J}^-
+i\hat{J}^+\hat{\sigma}_j^-\right)
-\hbar k\sin(k\hat{x}_j)\frac{\Gamma_\Delta}{2}\sum_{l=1}^N\cos(kx_l)
\left(\hat{\sigma}_j^+\hat{\sigma}_l^-+\hat{\sigma}_l^+\hat{\sigma}_j^-
-\beta_1 \hat{\sigma}_j^+\frac{d}{dt}\hat{\sigma}_l^--\beta_1^*
 \frac{d}{dt}\hat{\sigma}_l^+\hat{\sigma}_j^- \right)\\
&-\sin(k\hat{x}_j)\frac{\Gamma_C}{2}\sum_{l=1}^N
\frac{\eta}{2}\left[\sin(k\hat{x}_l),\hat{p}_l\right]_{+}\left(\hat{\sigma}_j^+\hat{\sigma}_l^-
+\hat{\sigma}_l^+\hat{\sigma}_j^-
+\beta_2\hat{\sigma}_j^+\hat{\sigma}_l^-+\beta_2^*
\hat{\sigma}_l^+\hat{\sigma}_j^-\right)+\hat{\mathcal{N}}_j\,, \label{fullforce}
\end{split}
\end{equation}
where $\left[\hat{A},\hat{B}\right]_{+}=\hat{A}\hat{B}+\hat{B}\hat{A}$ is the anticommutator and the coefficients are
\begin{equation}
\Gamma_\Delta=\frac{g^2\Delta/2}{\kappa^2/4+\Delta^2},\quad
\beta_1=\frac{\kappa}{\kappa^2/4+\Delta^2}+i\frac{\kappa^2/4
-\Delta^2}{\Delta(\kappa^2/4+\Delta^2)},\quad
\beta_2=i\frac{\kappa^2/4-\Delta^2}{\kappa\Delta},\quad
\eta=\frac{4\omega_\mathrm{r}\Delta}{\kappa^2/4+\Delta^2}\,.
\end{equation}
Here $\hat{\mathcal{N}}_j = \hat{\mathcal{N}}_j^{\text{cav}} + \hat{\mathcal{N}}_j^{\text{pump}} $ is the sum of the noise processes originating from the cavity output $\hat{\mathcal{N}}_j^{\text{cav}}$
and repumping $\hat{\mathcal{N}}_j^{\text{pump}}$. In the first line of equation (S\ref{fullforce}) we neglect $\beta_1$
because in the steady state superradiance regime it holds that $|\beta_1|
\langle\hat{\sigma}_j^+\frac{d}{dt}\hat{\sigma}_l^-\rangle\sim
\frac{w}{\kappa}\langle\hat{\sigma}_j^+\hat{\sigma}_l^-\rangle\ll
\langle\hat{\sigma}_j^+\hat{\sigma}_l^-\rangle$. This has also been checked numerically. Therefore we get
\begin{equation}\label{force+noise}
\begin{split}
\frac{d}{dt}\hat{p}_j &= \frac{d}{dt}\hat{p}_j^{0}+\hat{\mathcal{N}}_j\,,
\end{split}
\end{equation}
where we define the force without noise as
\begin{equation}
\begin{split}
\frac{d}{dt}\hat{p}_j^{0} &\approx \hbar
k\sin(k\hat{x}_j)\frac{\Gamma_C}{2}\left(-i\hat{\sigma}_j^+\hat{J}^-
+i\hat{J}^+\hat{\sigma}_j^-\right)
-\hbar k\sin(k\hat{x}_j)\frac{\Gamma_\Delta}{2}\sum_{l=1}^N\cos(kx_l)
\left(\hat{\sigma}_j^+\hat{\sigma}_l^-+\hat{\sigma}_l^+\hat{\sigma}_j^-
\right)\\
&-\sin(k\hat{x}_j)\frac{\Gamma_C}{2}\sum_{l=1}^N
\frac{\eta}{2}\left[\sin(k\hat{x}_l),\hat{p}_l\right]_{+}\left(\hat{\sigma}_j^+\hat{\sigma}_l^-
+\hat{\sigma}_l^+\hat{\sigma}_j^-
+\beta_2\hat{\sigma}_j^+\hat{\sigma}_l^-+\beta_2^*
\hat{\sigma}_l^+\hat{\sigma}_j^-\right)\,.
\end{split}
\end{equation}
We work at the detuning $\Delta=\kappa/2$ so that $\eta$ is maximized
and $\beta_2$ vanishes. As a result we obtain
\begin{equation}\label{force}
\frac{d}{dt}\hat{p}_j^0 \approx \hbar
k\sin(k\hat{x}_j)\frac{\Gamma_C}{2}\left(-i\hat{\sigma}_j^+\hat{J}^-
+i\hat{J}^+\hat{\sigma}_j^--
 \hat{\sigma}_j^+\hat{J}^--\hat{J}^+\hat{\sigma}_j^-\right)
-\sin(k\hat{x}_j)\frac{\Gamma_C}{2}\sum_{l=1}^N\frac{\eta}{2}\left[\sin(k\hat{x}_l),\hat{p}_l\right]_{+}\left(\hat{\sigma}_j^+\hat{\sigma}_l^-
+\hat{\sigma}_l^+\hat{\sigma}_j^-\right)\,.
\end{equation}
The first term on the RHS of Eq.~(S\ref{force}) represents forces originating from the adiabatic component of the cavity field, while the second term represents the frictional force arising from retardation effects.
The noise term  $\hat{\mathcal{N}}_j$ in equation \eqref{force+noise} gives rise to momentum diffusion due to
quantum noises associated with incoherent processes.
So we derive the equations of motion for the second moments of momenta,
\begin{equation}\label{smpq}
\begin{split}
\frac{d}{dt}\left\langle\hat{p}_j\hat{p}_l\right\rangle&
=\left\langle\hat{p}_j^0\frac{d\hat{p}_l^0}{dt}\right\rangle+
\left\langle\frac{d\hat{p}_j^0}{dt}\hat{p}_l^0\right\rangle
+\Gamma_C\hbar^2k^2\langle\sin(k\hat{x}_j)\sin(k\hat{x}_l)
\hat{\sigma}_j^+\hat{\sigma}_l^-\rangle+w\delta_{jl}
\hbar^2k'^2\overline{u^2}\langle\hat{\sigma}_j^-\hat{\sigma}_l^+\rangle\,,
\end{split}
\end{equation}
where $\delta_{jl}$ is the Kronecker delta, and $\overline{u^2}$ is the
second moment of the dipole radiation pattern, {\em i.e.},
\begin{equation}
\overline{u^2}=\int_{-1}^{1}du N(u)u^2 =\frac 2 5\,,
\end{equation}
where we have taken the dipole pattern $N(u)=\frac32|u|\sqrt{1-u^2}$.\\
We treat the external atomic motion classically under the assumption that the momentum width of the particles $\sqrt{\left \langle p^2\right\rangle}$ is larger than the single photon recoil $\hbar k$.
So we make the mapping $\langle\hat{p}_j\rangle\rightarrow p_j$
and $\langle\hat{x}_j\rangle\rightarrow x_j$. As a result this leads to
\begin{equation}\label{peom}
\frac{d}{dt}p_j=\frac{d}{dt}p_j^0+\xi_j^p\,,
\end{equation}
with
\begin{equation}
\frac{d}{dt}p_j^0=\hbar k\sin(kx_j)\Gamma_C\left(\mathrm{Im}[\langle\hat{\sigma}_j^+
\hat{J}^-\rangle]
-\mathrm{Re}[\langle\hat{\sigma}_j^+\hat{J}^-\rangle]\right)
-\sin(kx_j)\Gamma_C\sum_{l=1}^N\eta
\mathrm{Re}[\langle\hat{\sigma}_j^+\hat{\sigma}_l^-\rangle]
\sin(kx_l)p_l\,,
\end{equation}
where $\xi_j^p$ is the classical noise acting on the momentum of $j$-th atom and 
$\langle\xi_j^p(t)\xi_l^p(t')\rangle=D^{jl}\delta(t-t')$. The diffusion
matrix $D^{jl}$ can be computed by making quantum-classical correspondence
for the second moments. According to Eq.~(S\ref{peom}),
\begin{equation}\label{smpc}
\frac{d}{dt}\langle p_jp_l\rangle=
\left\langle p_j^0\frac{dp_l^0}{dt}\right\rangle+\left\langle\frac{dp_j^0}{dt}p_l^0\right\rangle+D^{jl}\,.
\end{equation}
We use symmetric ordering of quantum operators for
the quantum-classical correspondence, {\it i.e.}, 
$\frac{1}{2}\left\langle\left[\hat{p}_j,\frac{d\hat{p}_l}{dt}\right]_+\right\rangle
\rightarrow\left\langle p_j\frac{dp_l}{dt}\right\rangle$.
Matching Eq.~(S\ref{smpq}) and Eq.~(S\ref{smpc}), 
we get 
\begin{equation}
D^{jl}=\Gamma_C\hbar^2k^2\sin(kx_j)\sin(kx_l)\mathrm{Re}[\langle
\hat{\sigma}_l^+\hat{\sigma}_j^-\rangle]+w\delta_{jl}
\hbar^2k'^2\overline{u^2}\langle\hat{\sigma}_j^-\hat{\sigma}_l^+\rangle\,.
\end{equation}
Therefore, we could simulate the external motion of atoms
with Eq.~(S\ref{peom})  and the equation of motion for $x_j$
\begin{equation}\label{balistic}
\frac{d}{dt}x_j=\frac{p_j}{m}\,.
\end{equation}
The classical noises $\xi_j^p$ with diffusion matrix $D^{jl}$ make sure
that we have the right second order moments for momenta. \\ 
\section{III. Internal dynamics of atoms}
For the complete simulation of the atomic variables we also need to derive an equation for the internal degrees of freedom. In this section we will derive the equations of motions for the spins in which we drop third-order cumulants.
For the internal dynamics of atoms in a superradiant laser, it is
sufficient to keep the first order term in Eq.~(S\ref{ada}),
\begin{equation}\label{spad}
\hat{a}(t)\approx-i\frac{\Gamma_C}{g}\hat{J}^- -\frac{\Gamma_\Delta}{g}
\hat{J}^-+\hat{\mathcal{F}}(t)\,.
\end{equation}
Here, retardation effects are not included because they give rise to corrections that are of higher order and their contribution is negligible. This was also checked numerically.
The adiabatic elimination of the cavity field leads to an effective quantum master equation  
for the atomic spins only
\begin{equation}\label{efm}
\frac{d}{dt}\rho=\frac{1}{i\hbar}[\hat{H}_\mathrm{eff},\rho]
+\Gamma_C\mathcal{L}[\hat{J}^-]\rho+w
\sum_{j=1}^N\int_{-1}^{1} du N(u)
\mathcal{L}[\hat{\sigma}_j^+ e^{i\vec{k'}\cdot\vec{x}_j}]\rho\,,
\end{equation}
where the Hamiltonian
$\hat{H}_\mathrm{eff}=-\frac{\hbar\Gamma_\Delta}{2}\hat{J}^+\hat{J}^-$ describes
the coherent coupling between each pair of atoms, and the collective decay
[term $\Gamma_C\mathcal{L}[\hat{J}^-]$ in Eq.~(S\ref{efm})] leads to dissipative coupling.
We want to emphasize that this atomic master equation is not sufficient for the external degrees of freedom, 
which are treated in section II separately, and for which retardation effects are not negligible.

The spin degrees of freedom of atoms scale exponentially with the number of 
atoms. To solve Eq.~(S\ref{efm}), we thus use a semiclassical approximation
that is applicable to large atom numbers in the steady-state
superradiance~\cite{s1,s2}. Cumulants for the expectation values of spin
operators are expanded to second order. Because of the U(1) symmetry, $\langle
\hat{\sigma}_j^\pm\rangle=0$.
Therefore, all nonzero observables are expanded in terms
of $\langle\hat{\sigma}_j^+\hat{\sigma}_j^-\rangle$ and 
$\langle\hat{\sigma}_j^+\hat{\sigma}_l^-\rangle$~($j\ne l$). Their equations of motion
can then be found from the effective master equation,
\begin{equation}\label{cumu}
\begin{split}
\frac{d}{dt}\langle\hat{\sigma}_j^+\hat{\sigma}_j^-\rangle&=
w(1-\langle\hat{\sigma}_j^+\hat{\sigma}_j^-\rangle)
-\frac{1}{2}(\Gamma_C+i\Gamma_\Delta)\cos(k\hat{x}_j)
\langle\hat{J}^+\hat{\sigma}_j^-\rangle
-\frac{1}{2}(\Gamma_C-i\Gamma_\Delta)\cos(k\hat{x}_j)
\langle\hat{\sigma}_j^+\hat{J}^-\rangle,\\
\frac{d}{dt}\langle\hat{\sigma}_j^+\hat{\sigma}_l^-\rangle&=
-w\langle\hat{\sigma}_j^+
\hat{\sigma}_l^-\rangle+\frac{1}{2}(\Gamma_C+i\Gamma_\Delta)\cos(k\hat{x}_j)
\langle\hat{J}^+\hat{\sigma}_l^-\hat{\sigma}_j^z\rangle
+\frac{1}{2}(\Gamma_C-i\Gamma_\Delta)\cos(k\hat{x}_l)
\langle\hat{\sigma}_l^z\hat{\sigma}_j^+\hat{J}^-\rangle\\
&\approx-\left(w+(\Gamma_C+i\Gamma_\Delta)\cos^2(k\hat{x}_j)
\langle\hat{\sigma}_j^+\hat{\sigma}_j^-\rangle
+(\Gamma_C-i\Gamma_\Delta)\cos^2(k\hat{x}_l)
\langle\hat{\sigma}_l^+\hat{\sigma}_l^-\rangle\right)
\langle\hat{\sigma}_j^+\hat{\sigma}_l^-\rangle\\
&\quad+\frac{1}{2}(\Gamma_C+i\Gamma_\Delta)\cos(k\hat{x}_j)
(2\langle\hat{\sigma}_j^+\hat{\sigma}_j^-\rangle-1)
\langle\hat{J}^+\hat{\sigma}_l^-\rangle
+\frac{1}{2}(\Gamma_C-i\Gamma_\Delta)\cos(k\hat{x}_l)
(2\langle\hat{\sigma}_l^+\hat{\sigma}_l^-\rangle-1)
\langle\hat{\sigma}_j^+\hat{J}^-\rangle,
\end{split}
\end{equation}
describing the population inversion and spin-spin correlation respectively.
In deriving Eq.~(S\ref{cumu}), we have dropped the third-order cumulants.
In the simulations we integrate \eqref{peom}, \eqref{balistic} and \eqref{cumu} simultaneously.


\begin{thebibliography}{PRL}
\bibitem{note}This requires $\kappa\gg\sqrt{N\bar{n}}g$, $w$, and 
$k\sqrt{\langle p^2\rangle}/m$, where $\bar{n}$ is the mean photon number
in the cavity.
\end{thebibliography}
\end{document}